\PassOptionsToPackage{prologue,table}{xcolor}
\documentclass[sigconf]{acmart}
\acmConference[MSR 2024]{21st International Conference on Mining Software Repositories}{April 2024}{Lisbon, Portugal}

\AtBeginDocument{%
  }
\newcommand*\smarthspace[1]{%
    \ifhmode
        \hspace{#1}%
    \fi
}

\usepackage[table]{xcolor}
\usepackage{graphicx, subfigure}
\usepackage{caption}
\usepackage{makecell}
\usepackage{soul}
\usepackage{listings}
\usepackage{multirow}
\usepackage{pifont}
\usepackage{xspace}
\usepackage{epigraph} 
\usepackage[flushleft]{threeparttable}
\usepackage{url}
\usepackage{hyperref}
\definecolor{backcolour}{rgb}{0.95,0.95,0.92}
\usepackage{algorithm2e}
\UseRawInputEncoding
\usepackage[colorinlistoftodos]{todonotes}
\usepackage{listings}
\usepackage{comment}
\usepackage{fontawesome}
\usepackage{balance}
\usepackage{CJKutf8}

\definecolor{celestialblue}{rgb}{0.29, 0.59, 0.82}
\definecolor{awesome}{rgb}{0.0, 0.2, 0.6}
\definecolor{coolblack}{rgb}{0.0, 0.18, 0.39}
\definecolor{maroon}{cmyk}{0, 0.87, 0.68, 0.32}
\definecolor{halfgray}{gray}{0.55}
\definecolor{ipython_frame}{RGB}{207, 207, 207}
\definecolor{ipython_bg}{RGB}{247, 247, 247}
\definecolor{ipython_red}{RGB}{186, 33, 33}
\definecolor{ipython_green}{RGB}{0, 128, 0}
\definecolor{ipython_cyan}{RGB}{64, 128, 128}
\definecolor{ipython_purple}{RGB}{170, 34, 255}
\usepackage[strict]{changepage}
  \definecolor{ABlue}{HTML}{127bca}
 \definecolor{LHScolor}{HTML}{555555}
\usepackage{framed}

\definecolor{mygreen}{rgb}{0,0.6,0}
\definecolor{mygray}{rgb}{0.5,0.5,0.5}
\definecolor{mymauve}{rgb}{0.58,0,0.82}
\definecolor{ceruleanblue}{rgb}{0.16, 0.32, 0.75}

\definecolor{formalshade}{rgb}{0.95,0.96,0.96}
\definecolor{side}{rgb}{0.0,0.2,0.6}

\newenvironment{formal}{%
  \MakeFramed{\advance\hsize-\width\FrameRestore}%
  \noindent\hspace{-4.55pt}
  \begin{adjustwidth}{}{7pt}%
  \vspace{2pt}\vspace{2pt}%
}
{%
  \vspace{2pt}\end{adjustwidth}\endMakeFramed%
}
\definecolor{gray(x11gray)}{rgb}{0.75, 0.75, 0.75}

\lstdefinelanguage{python}{
    morekeywords={access,and,break,class,continue,def,del,elif,else,except,exec,finally,for,from,global,if,import,in,is,lambda,not,or,pass,print,raise,return,try,while},
    morekeywords=[2]{abs,all,any,basestring,bin,bool,bytearray,callable,chr,classmethod,cmp,compile,complex,delattr,dict,dir,divmod,enumerate,eval,execfile,file,filter,float,format,frozenset,getattr,globals,hasattr,hash,help,hex,id,input,int,isinstance,issubclass,iter,len,list,locals,long,map,max,memoryview,min,next,object,oct,open,ord,pow,property,range,raw_input,reduce,reload,repr,reversed,round,set,setattr,slice,sorted,staticmethod,str,sum,super,tuple,type,unichr,unicode,vars,xrange,zip,apply,buffer,coerce,intern},
    sensitive=true,
    morecomment=[l]\#,
    morestring=[b]',
    morestring=[b]",
    morestring=[s]{'''}{'''},
    morestring=[s]{"""}{"""},
    morestring=[s]{r'}{'},
    morestring=[s]{r"}{"},
    morestring=[s]{r'''}{'''},
    morestring=[s]{r"""}{"""},
    morestring=[s]{u'}{'},
    morestring=[s]{u"}{"},
    morestring=[s]{u'''}{'''},
    morestring=[s]{u"""}{"""},
    literate=
    {á}{{\'a}}1 {é}{{\'e}}1 {í}{{\'i}}1 {ó}{{\'o}}1 {ú}{{\'u}}1
    {Á}{{\'A}}1 {É}{{\'E}}1 {Í}{{\'I}}1 {Ó}{{\'O}}1 {Ú}{{\'U}}1
    {à}{{\`a}}1 {è}{{\`e}}1 {ì}{{\`i}}1 {ò}{{\`o}}1 {ù}{{\`u}}1
    {À}{{\`A}}1 {È}{{\'E}}1 {Ì}{{\`I}}1 {Ò}{{\`O}}1 {Ù}{{\`U}}1
    {ä}{{\"a}}1 {ë}{{\"e}}1 {ï}{{\"i}}1 {ö}{{\"o}}1 {ü}{{\"u}}1
    {Ä}{{\"A}}1 {Ë}{{\"E}}1 {Ï}{{\"I}}1 {Ö}{{\"O}}1 {Ü}{{\"U}}1
    {â}{{\^a}}1 {ê}{{\^e}}1 {î}{{\^i}}1 {ô}{{\^o}}1 {û}{{\^u}}1
    {Â}{{\^A}}1 {Ê}{{\^E}}1 {Î}{{\^I}}1 {Ô}{{\^O}}1 {Û}{{\^U}}1
    {œ}{{\oe}}1 {Œ}{{\OE}}1 {æ}{{\ae}}1 {Æ}{{\AE}}1 {ß}{{\ss}}1
    {ç}{{\c c}}1 {Ç}{{\c C}}1 {ø}{{\o}}1 {å}{{\r a}}1 {Å}{{\r A}}1
    {€}{{\EUR}}1 {£}{{\pounds}}1
    {^}{{{\color{ipython_purple}\^{}}}}1
    {=}{{{\color{ipython_purple}=}}}1
    {+}{{{\color{ipython_purple}+}}}1
    {*}{{{\color{ipython_purple}$^\ast$}}}1
    {/}{{{\color{ipython_purple}/}}}1
    {+=}{{{+=}}}1
    {-=}{{{-=}}}1
    {*=}{{{$^\ast$=}}}1
    {/=}{{{/=}}}1,
    literate=
    *{-}{{{\color{ipython_purple}-}}}1
     {?}{{{\color{ipython_purple}?}}}1,
    identifierstyle=\color{black}\ttfamily,
    commentstyle=\color{ipython_cyan}\ttfamily,
    stringstyle=\color{ipython_red}\ttfamily,
    keepspaces=true,
    showspaces=false,
    showstringspaces=false,
    rulecolor=\color{ipython_frame},
    numberstyle=\tiny\color{halfgray},
    backgroundcolor=\color{ipython_bg},
    basicstyle=\scriptsize,
    keywordstyle=\color{ipython_green}\ttfamily,
}

\definecolor{chestnut}{rgb}{0.8, 0.36, 0.36}

\definecolor{chestnut}{rgb}{0.8, 0.36, 0.36}

\usepackage[skins,breakable]{tcolorbox}

\definecolor{Large}{HTML}{696969}
\definecolor{Negligible}{HTML}{D3D3D3}
\definecolor{Medium}{HTML}{808080}
\definecolor{Small}{HTML}{A9A9A9}

\begin{document}

\title{Exploring the Effect of Multiple Natural Languages on Code Suggestion Using GitHub Copilot}

\author{Kei Koyanagi}
\email{koyanagi@posl.ait.kyushu-u.ac.jp}
\author{Dong Wang}
\email{d.wang@ait.kyushu-u.ac.jp}
\author{Kotaro Noguchi}
\email{noguchi@posl.ait.kyushu-u.ac.jp}
\author{Masanari Kondo}
\email{kondo@ait.kyushu-u.ac.jp}
\affiliation{%
  \institution{Kyushu University}
  \country{Japan}
}
\author{Alexander Serebrenik}
\email{a.serebrenik@tue.nl}
\affiliation{%
  \institution{Eindhoven University of Technology}
  \country{The Netherlands}
}
\author{Yasutaka Kamei}
\email{kamei@ait.kyushu-u.ac.jp}
\author{Naoyasu Ubayashi}
\email{ubayashi@ait.kyushu-u.ac.jp}
\affiliation{%
  \institution{Kyushu University}
  \country{Japan}
}

\begin{abstract}
GitHub Copilot is an AI-enabled tool that automates program synthesis. It has gained significant attention since its launch in 2021.
Recent studies have extensively examined Copilot's capabilities in various programming tasks, as well as its security issues.
However, little is known about the effect of different natural languages on code suggestion. 
Natural language is considered a social bias in the field of NLP, and this bias could impact the diversity of software engineering.
To address this gap, we conducted an empirical study to investigate the effect of three popular natural languages (English, Japanese, and Chinese) on Copilot. 
We used 756 questions of varying difficulty levels from AtCoder contests for evaluation purposes.
The results highlight that the capability varies across natural languages, with Chinese achieving the worst performance.
Furthermore, regardless of the type of natural language, the performance decreases significantly as the difficulty of questions increases.
Our work represents the initial step in comprehending the significance of natural languages in Copilot's capability and introduces promising opportunities for future endeavors.
\end{abstract}



\keywords{Code Suggestion, GitHub Copilot, Empirical Study}

\maketitle

\section{Introduction}
In recent years, the expansion of IT demand has led to the use of various support tools, such as task management tools and project management tools, to improve development efficiency.
One of these tools is GitHub Copilot, which was introduced by GitHub and OpenAI in June 2022~\cite{Copilot}. 
Copilot is a code suggestion tool powered by a large-scale language model. 
It suggests code snippets and libraries in different programming languages to developers based on comments that describe specifications and the code being written.
As a result, developers could save time by not starting from scratch and further reduce development costs.

It is well-known that the output of large language models can vary significantly depending on the input~\cite{lu2021fantastically}.
Several studies have been conducted to empirically examine the impact of input on the accuracy of suggested code in terms of Copilot.
For example, \citet{yetistiren2022assessing} reported that the utilization of proper
explanations of the given problem is important in terms of acquiring correct and valid code. 
\citet{nguyen2022empirical} evaluated Copilot's performance by running LeetCode's provided tests. The results showed that Copilot's performance varies across different languages, with Java having the highest correctness.
\citet{mastropaolo2023robustness} revealed that paraphrasing the input description results in different code suggestions with substantial variations, indicating the central role played by the model's input.

To reduce barriers for non-native English speakers and assist developers of all backgrounds, Copilot supports not only English but also other languages such as Chinese and Japanese.
In addition to languages of specific domains, the performance of large language models can also be influenced by the characteristics of natural languages~\cite{kwiatkowski2019natural, wang-etal-2023-mconala}.
Recent studies have proposed a benchmark that spans multiple natural languages in order to address the technology gap in code suggestion.
For example, \citet{wang-etal-2023-mconala} developed a multilingual dataset namely MCoNala, annotating a total of 896 NL-Code pairs in Spanish, Japanese, and Russian.
However, little is known about the effect of diverse natural languages on the Copilot tool. 
We hypothesize that variations in natural language input could also substantially affect the performance of Copilot.

To minimize the potential bias caused by natural languages in the future, we conduct an empirical study on Copilot to explore the effect of multiple natural languages on code suggestion.  
Specifically, we use 756 AtCoder (a programming contest) questions from 189 contests to create queries for Copilot in English, Japanese, and Chinese.
We then evaluate the correctness of the corresponding  Copilot suggestions against the test cases provided by AtCoder. 

The preliminary results reveal that natural languages do have an effect on the correctness of the code suggestions when using Copilot. 
We find that in the context of AtCoder questions, Japanese has the highest level of correctness, followed by English, while Chinese performs the worst.
More specifically, there is a difference ranging from 2.6\% to 11.5\% observed between Chinese and English across four levels of question difficulty.
In addition, the correctness tends to decrease dramatically with the increase of difficulty for all studied natural languages.

In summary, the contributions of this work are as follows: (1) we are the first work to investigate the ability of Copilot to suggest code, taking into account the distinction of natural languages; (2) the results emphasize the significance of addressing bias in natural languages when using generative AI techniques for programming tasks, in addition to efforts aimed
at addressing political~\cite{liu2021mitigating}, sentiment~\cite{huang2019reducing}, and gender~\cite{treude2023she} aspects; (3) Our work opens up several future research directions.

\section{Related Work}
\smallskip
\noindent
\textbf{Studies on GitHub Copilot.} 
Several studies have focused on the securities of Copilot.
\citet{pearce2022asleep} examined the security issues in code suggested based on queries created from the 25 top CWE vulnerabilities with a total of 89 scenarios, with approximately 40\% of them being vulnerable.
\citet{asare2023github} demonstrated that Copilot, despite performing differently across various vulnerability types, is not as bad
as human developers at introducing vulnerabilities.
At the same time, another group of studies have been conducted to evaluate the various capabilities of Copilot.
To name a few, \citet{nguyen2022empirical} validated the correctness of 132 LeetCode questions across four different programming languages.
\citet{mastropaolo2023robustness} analyzed whether different but semantically equivalent natural language
descriptions result in the same recommended function.
\citet{dakhel2023github} investigated the quality of the code by Copilot and suggested that Copilot can become an asset for experts, but a liability for novice developers.
\citet{sobania2022choose} compared the Copilot and Genetic Programming on standard program synthesis benchmark problems.
\citet{al2022readable} reported that the suggested code is comparable in complexity and readability to code written by human pair programmers.

\smallskip
\noindent
\textbf{Studies on Software Engineering Techniques for Multilingual Projects.}
Several studies have been conducted to investigate the impact of natural languages on software engineering techniques, such as traceability link recovery, bug localization, and question retrieval~\cite{lin2022EMSE,hayes2011TSFSE,xia2014ICPC,pawelka2015ICSME,xu2016MSR,chen2016ASE}. 
For instance, \citet{lin2022EMSE} evaluated different approaches for traceability link recovery in bilingual projects, including traditional IR-based, language-model based, and deep learning model-based approaches. They conducted their evaluation on 14 English-Chinese projects and three projects in other languages, such as Japanese. The results showed that the deep learning model-based approach outperformed the other approaches, particularly in large-scale projects in this bilingual setting.

\smallskip
\noindent
\textbf{Studies on Social Bias in Language Models.} Several studies have been carried out toward understanding and mitigating social biases in language models~\cite{huang2019reducing, liang2021towards, wan2023biasasker}.
Regarding gender bias, \citet{treude2023she} examined the extent to which 56 tasks related to software development are affected by implicit gender bias embedded in large language models.
Their findings revealed
a clear pattern of gender bias.
Regarding religion bias, \citet{abid2021persistent} investigated the anti-Muslim bias, demonstrating that it appears consistently and creatively in different uses of the model and that it is severe even compared to biases about other religious groups.
Regarding political bias, \citet{liu2021mitigating} described metrics for measuring political bias and proposed a reinforcement learning framework for mitigating such biases in the generated text.
Regarding sentiment bias, \citet{huang2019reducing} quantified sentiment bias by adopting individual and
group fairness metrics from the fair machine
learning literature.

\textit{In contrast to previous work, we present a novel perspective to examine the Copilot capability within the context of diverse natural languages on code suggestion.
This would offer valuable insights into optimizing the usage of Copilot and complement the knowledge of social bias in large language models.}

\section{Study Design}
In this section, we describe the study design, including the research question, data collection, and code suggestion and its evaluation.

\subsection{Research Question}

\textbf{RQ - \ul{How does the input of different natural languages affect the performance of Copilot?}}
We aim to study the effectiveness of natural languages (i.e., English, Japanese, and Chinese) in the code suggestions synthesized by the Copilot. 
Answering this research question will shed light on the usage of Copilot for developers from different backgrounds in practice.

\subsection{Data Collection}
To answer our research question, we collected the programming questions from the AtCoder.
AtCoder is one of the largest programming contest sites, catering to individuals ranging from beginners to experts. It has been extensively studied in previous works analyzing programming~\cite{mathew2021cross, ahmad2021avatar}.
Hence, it is a suitable starting point for examining the impact on code suggestion.

We used the dataset that was publicly available at the end of March 2023, consisting of 1,624 questions from 287 contests.
Each contest has up to eight levels of difficulty, labeled A, B, C, D, E, F, G, and Ex (H). The higher the letter's position in the alphabet, the more difficult the question.
Some contests did not contain all eight levels.
To minimize potential bias in our validation, we established the following two criteria for selecting the questions:
\begin{itemize}
    \item We focus on levels A to D since questions E and higher require complex processing that involves multiple algorithms.
    \item We excluded the contests where the test cases were not provided.
\end{itemize} 
After applying the criteria above, we collected a total of 756 questions (189 per level) from 189 contests (from the 99th contest to the 287th contest).
We chose the 99th because the following contests consistently feature four levels of difficulty.
Note that AtCoder only offers questions in English and Japanese.
Therefore, we need to translate the questions into Chinese. Chinese is a representative language in this context study~\cite{lin2022EMSE,xia2014ICPC,xu2016MSR,chen2016ASE} and serves as the comparative language for our goal. 
To ensure accuracy, the second author, a native Chinese speaker with over five years of experience in software engineering research inspected all 756 questions translated by \texttt{DeepL API}\footnote{\url{https://www.deepl.com/docs-api}} and made significant modifications.
A similar method has also been adopted by Wan et al.~\cite{wan2023biasasker}.  
Below we present four examples of question descriptions with levels A to D from the 212th contest\footnote{\url{https://atcoder.jp/contests/abc212/tasks}} for a better understanding of code suggestion tasks. 

\smallskip
\noindent
The question with \textbf{A} level:
\begin{formal}
Takahashi melted and mixed $A$ grams of gold and $B$ grams of silver ($0 \le A, B, 0 < A+B$) to produce new metal. What metal did he produce: pure gold, pure silver, or an alloy? Formally, the product is called as follows. Pure gold, if $0 < A$ and $B=0$. Pure silver, if $A=0$ and $0 < B$. An alloy, if $0 < A$ and $0 < B$.
\end{formal}

\smallskip
\noindent
The question with \textbf{B} level:
\begin{formal}
You are given a 4-digit PIN: $X_1X_2X_3X_4$, which may begin with a 0.
The PIN is said to be weak when it satisfies one of the following conditions:
All of the four digits are the same.
For each integer $i$ such that $1\le i\le 3$, $X_{i+1}$ follows $X_i$. Here, $j+1$ follows $j$ for each $0\le j\le 8$, and 0 follows 9.
If the given PIN is weak, print Weak; otherwise, print Strong.
\end{formal}

\smallskip
\noindent
The question with \textbf{C} level:
\begin{formal}
You are given two sequences: $A=(A_1,A_2, ... ,A_N)$ consisting of $N$ positive integers, and $B=(B_1, ... ,B_M)$ consisting of $M$ positive integers.
Find the minimum difference of an element of $A$ and an element of $B$, that is,  $\underset{1\le i\le N}{\rm{min}} \underset{1\le j\le M}{\rm{min}} | A_i-B_j|$.
\end{formal}

\smallskip
\noindent
The question with \textbf{D} level:
\begin{formal}
Takahashi has many balls, on which nothing is written, and one bag.
Initially, the bag is empty. Takahashi will do $Q$ operations, each of which is of one of the following three types.
Type 1: Write an integer $X_i$ on a blank ball and put it in the bag.
Type 2: For each ball in the bag, replace the integer written on it with that integer plus $X_i$.
Type 3: Pick up the ball with the smallest integer in the bag (if there are multiple such balls, pick up one of them). Record the integer written on this ball and throw it away.
For each $1\le i\le Q$, you are given the type $P_i$ of the $i$-th operation and the value of $X_i$ if the operation is of Type 1 or 2. Print the integers recorded in the operations of Type 3 in order.
\end{formal}

\subsection{Code Suggestion and Evaluation}
\smallskip
\noindent
\textbf{Code Suggestion.}
We invoked Copilot to suggest the code using a dataset of 756 questions from three different natural languages. 
In this study, our focus is on code written in Python, which is currently a widely used programming language.
During the input, in addition to the question description, we provided complementary information to make the query more precise. This included question constraints, the format of input and output, as well as examples of input and output.
For example, given the aforementioned question with \textbf{A} level, the complementary information is:
\begin{center}
\begin{tabular}{c}
\begin{lstlisting}[language=Python,
caption={}, basicstyle=\footnotesize, linewidth=.8\columnwidth, 
label=code:userhome]
[Constraint] 0 <= A, B <= 100 and 1 <= A+B           and A,B are integers
[Input] A B
[Output] Gold or Silver or Alloy
[Input Example] 50 50
[Output Example] Alloy
\end{lstlisting}
\end{tabular}
\end{center}

Based on the input, Copilot suggested \textit{x} code snippets, 0$\le$\textit{x}$\le$10.
Unlike the prior work~\cite{nguyen2022empirical} where they manually invoked Copilot for each query, to mitigate the human bias, we established an environment using Keyboard Maestro\footnote{\url{https://www.keyboardmaestro.com/main/}} to automatically manipulate the input/suggestion.
Note that the process from input to output is executed five times for every queried question. 
The suggested code by Copilot may vary each time, hence we take into account the potential randomness bias in our subsequent evaluation.

\smallskip
\noindent
\textbf{Evaluation.}
We evaluated the suggested code using the corresponding test cases provided for each question. Specifically, we examined whether the output of the suggested code matches the expected output of the test case.
We used \textit{Accuracy} to measure the correctness of the suggested code, which indicates the percentage of code that passed all test cases out of all suggested code.
To statistically confirm the significant differences among the studied natural languages at each difficulty level of questions, we perform the one-way ANOVA test~\citep{fisher1970statistical}.

\section{Results}
Tables \ref{En_accuracy}, \ref{Ja_accuracy} and \ref{Zh_accuracy} show the accuracy of suggested code for five rounds under the setting of English, Japanese, and Chinese,  respectively.
Each row displays the proportion of suggested code snippets that passed all test cases in each round.  
Note that the number of suggested snippets varies depending on the questions.

\ul{\textit{When the query is written in Chinese, Copilot is less likely to suggest the correct code}}.
As shown in Table \ref{Zh_accuracy}, we observe that the accuracy of Chinese is relatively lower than that of English and Japanese for all levels of questions, which indicates that the suggested code has a lower chance of passing the provided test cases.
For example, the accuracy of the \textbf{A}, \textbf{B}, \textbf{C}, \textbf{D} level, varies from 51.5\% to 52.7\%, 39.7\% to 43.1, 21.9\% to 23.2\%, and 7.6\% to 8.4\%.
However, as shown in Table \ref{En_accuracy}, when the query is written in English, the accuracy for the four levels, ranges from 60.4\% to 65.4\%, 50.5\% to 51.6\%, 26.6\% to 31.3\%, and 9.3\% to 12.0\%, respectively.
On the other hand, we find that the Japanese achieved the best performance comparatively.
Specifically, the accuracy of the \textbf{A}, \textbf{B}, \textbf{C}, \textbf{D} level varies from 67.4\% to 68.7\%, 52.8\% to 57.6\%, 28.4\% to 33.1\%, and 10.0\% to 12.8\%.
Moreover, the statistical tests indicate that there are significant differences among the three languages at each difficulty level.
The p-values of the statistical tests for the A, B, C, and D levels among the three languages are 9.924e-10, 1.258e-10, 7.661e-10, and 0.000256, respectively.

  \begin{table}[]
    \centering
    \caption{Accuracy of Suggested Code--English}
    \label{En_accuracy}
    \scalebox{0.73}{
\begin{tabular}{c|cc|cc|cc|cc}
\toprule
       & \multicolumn{2}{c|}{A}                 & \multicolumn{2}{c|}{B}                 & \multicolumn{2}{c|}{C}                 & \multicolumn{2}{c}{D}                  \\ \midrule
1st    & \multicolumn{1}{c|}{63.7\%} & 864/1357 & \multicolumn{1}{c|}{51.0\%} & 860/1686 & \multicolumn{1}{c|}{28.3\%} & 482/1702 & \multicolumn{1}{c|}{10.4\%} & 172/1657 \\ \midrule
2nd    & \multicolumn{1}{c|}{65.4\%} & 878/1343 & \multicolumn{1}{c|}{50.5\%} & 858/1698 & \multicolumn{1}{c|}{26.7\%} & 471/1761 & \multicolumn{1}{c|}{10.0\%} & 182/1813 \\ \midrule
3rd    & \multicolumn{1}{c|}{63.6\%} & 856/1346 & \multicolumn{1}{c|}{50.9\%} & 811/1594 & \multicolumn{1}{c|}{26.6\%} & 438/1648 & \multicolumn{1}{c|}{9.3\%}  & 151/1622 \\ \midrule
4th    & \multicolumn{1}{c|}{60.4\%} & 819/1355 & \multicolumn{1}{c|}{51.6\%} & 842/1631 & \multicolumn{1}{c|}{31.3\%} & 499/1596 & \multicolumn{1}{c|}{11.0\%} & 167/1524 \\ \midrule
5th    & \multicolumn{1}{c|}{60.4\%} & 831/1375 & \multicolumn{1}{c|}{50.5\%} & 857/1698 & \multicolumn{1}{c|}{31.1\%} & 552/1774 & \multicolumn{1}{c|}{12.0\%} & 217/1812 \\ \midrule
Median & \multicolumn{2}{c|}{63.6\%}            & \multicolumn{2}{c|}{50.9\%}            & \multicolumn{2}{c|}{28.3\%}            & \multicolumn{2}{c}{10.4\%}             \\ \bottomrule
\end{tabular}
    }
  \end{table}

  \begin{table}[]
    \centering
    \caption{Accuracy of Suggested Code--Japanese}
    \label{Ja_accuracy}
    \scalebox{0.73}{
\begin{tabular}{c|cc|cc|cc|cc}
\toprule
       & \multicolumn{2}{c|}{A}                 & \multicolumn{2}{c|}{B}                 & \multicolumn{2}{c|}{C}                 & \multicolumn{2}{c}{D}                  \\ \midrule
1st    & \multicolumn{1}{c|}{68.7\%} & 857/1248 & \multicolumn{1}{c|}{52.8\%} & 872/1651 & \multicolumn{1}{c|}{28.9\%} & 482/1668 & \multicolumn{1}{c|}{10.0\%} & 170/1697 \\ \midrule
2nd    & \multicolumn{1}{c|}{67.4\%} & 840/1246 & \multicolumn{1}{c|}{52.9\%} & 824/1559 & \multicolumn{1}{c|}{28.4\%} & 454/1597 & \multicolumn{1}{c|}{10.2\%} & 155/1526 \\ \midrule
3rd    & \multicolumn{1}{c|}{68.1\%} & 842/1237 & \multicolumn{1}{c|}{54.3\%} & 879/1619 & \multicolumn{1}{c|}{28.5\%} & 449/1577 & \multicolumn{1}{c|}{11.0\%} & 153/1391 \\ \midrule
4th    & \multicolumn{1}{c|}{68.4\%} & 860/1258 & \multicolumn{1}{c|}{57.6\%} & 872/1515 & \multicolumn{1}{c|}{33.1\%} & 439/1328 & \multicolumn{1}{c|}{12.8\%} & 125/974  \\ \midrule
5th    & \multicolumn{1}{c|}{68.0\%} & 849/1249 & \multicolumn{1}{c|}{54.5\%} & 922/1693 & \multicolumn{1}{c|}{32.7\%} & 592/1813 & \multicolumn{1}{c|}{11.4\%} & 208/1832 \\ \midrule
Median & \multicolumn{2}{c|}{68.1\%}            & \multicolumn{2}{c|}{54.3\%}            & \multicolumn{2}{c|}{28.9\%}            & \multicolumn{2}{c}{11.0\%}             \\ \bottomrule
\end{tabular}
    }
  \end{table}

  \begin{table}[]
    \centering
    \caption{Accuracy of Suggested Code--Chinese}
    \label{Zh_accuracy}
    \scalebox{0.73}{
\begin{tabular}{c|cc|cc|cc|cc}
\toprule
                            & \multicolumn{2}{c|}{A}                 & \multicolumn{2}{c|}{B}                 & \multicolumn{2}{c|}{C}                 & \multicolumn{2}{c}{D}                 \\ \midrule
1st                         & \multicolumn{1}{c|}{52.7\%} & 739/1402 & \multicolumn{1}{c|}{40.1\%} & 705/1757 & \multicolumn{1}{c|}{22.7\%} & 405/1787 & \multicolumn{1}{c|}{7.7\%} & 136/1766 \\ \midrule
2nd                         & \multicolumn{1}{c|}{52.5\%} & 728/1386 & \multicolumn{1}{c|}{43.1\%} & 734/1703 & \multicolumn{1}{c|}{21.9\%} & 376/1718 & \multicolumn{1}{c|}{7.8\%} & 134/1715 \\ \midrule
3rd                         & \multicolumn{1}{c|}{52.1\%} & 737/1415 & \multicolumn{1}{c|}{39.7\%} & 687/1730 & \multicolumn{1}{c|}{22.7\%} & 405/1788 & \multicolumn{1}{c|}{8.1\%} & 143/1764 \\ \midrule
4th                         & \multicolumn{1}{c|}{51.8\%} & 732/1412 & \multicolumn{1}{c|}{41.5\%} & 717/1728 & \multicolumn{1}{c|}{23.2\%} & 413/1783 & \multicolumn{1}{c|}{8.4\%} & 146/1748 \\ \midrule
5th                         & \multicolumn{1}{c|}{51.5\%} & 739/1436 & \multicolumn{1}{c|}{41.9\%} & 728/1738 & \multicolumn{1}{c|}{22.0\%} & 395/1798 & \multicolumn{1}{c|}{7.6\%} & 132/1744 \\ \midrule
\multicolumn{1}{l|}{Median} & \multicolumn{2}{c|}{52.1\%}            & \multicolumn{2}{c|}{41.5\%}            & \multicolumn{2}{c|}{22.7\%}            & \multicolumn{2}{c}{7.8\%}             \\ \bottomrule
\end{tabular}
    }
  \end{table}

\ul{\textit{The accuracy varies as the number of execution rounds changes}}.
Our repetitive experiment suggests that to some extent there is a bias towards randomness in Copilot for all studied natural languages.
For instance, as we can see from Table \ref{En_accuracy}, for English, there are performance differences of up to 5.0\% for \textbf{A} level, 1.1\% for \textbf{B} level, 4.7\% for \textbf{C} level, and 2.7\% for \textbf{D} level.
Similarly, in the context of Japanese, the performance differences are 1.3\%, 4.8\%, 4.7\%, and 2.8\% for the four levels, separately (Table~\ref{Ja_accuracy}).

In addition, regardless of the type of natural language, the performance of Copilot decreases significantly as the difficulty of questions increases.
As shown in the three tables, the accuracy continuously decreases to approximately 10\% or less from \textbf{A} level to \textbf{D} level (e.g., within Japanese, 68.7\% for \textbf{A} while 10.0 for \textbf{D}).  


\section{Discussion}

\ul{\textit{(I) Low performance when input is Chinese}}. The results reveal that Copilot performed the worst in the Chinese setting compared to the other two languages.
This finding is not surprising, considering that AtCoder only offers questions in Japanese and English. 
Hence, the number of codes with Chinese comments used during the training of the large language models is likely to be limited.
One potential future direction is to investigate the effectiveness of Copilot using a dataset from a Chinese-based programming contest as the baseline.

To shed light on the reasons for the low accuracy, we manually inspected some instances where the suggested code passed all the test cases for English and Japanese but failed for Chinese.
Based on our inspection, we have found that Copilot has a tendency to suggest incorrect code in situations where the correct code should handle multiple conditions, complex conditions, or string outputs.
The below example displays one of the suggested incorrect code snippets for Chinese in question 212-A. 
\begin{center}
\begin{tabular}{c}
\begin{lstlisting}[language=Python,
caption={}, linewidth=.7\columnwidth, basicstyle=\footnotesize, 
label=code:userhome]
def main():
    A, B = map(int, input().split())
    if A == 0:
        print("Silver")
    elif B == 0:
        print("Gold")
\end{lstlisting}
\end{tabular}
\end{center}
The correct code should handle three conditions: Gold, Silver, and Alloy, based on the combination of variables A and B. However, this code only has two conditions.
This tendency may be because the output format of AtCoder strings uses Roman characters or English words, and the Chinese dataset used as input included Roman characters or English words. These characters may not have been recognized as symbols.

\ul{\textit{(II) Relationship between the number of suggested code and the correctness}}.
When analyzing Tables 1, 2, and 3 and focusing on maximum accuracy, if there is a difference of 3\% or more from the lowest accuracy across five rounds within the same language and difficulty level, the total number of suggested code snippets will be the lowest. This result suggests that increasing the maximum number of suggested code snippets may decrease accuracy in comparison to suggesting a smaller number.

In terms of the number of code snippets suggested for each difficulty level, as shown in the tables, we notice that level \textbf{A} had the lowest number of code snippets suggested compared to levels \textbf{B}, \textbf{C}, and \textbf{D}. 
This might be because Copilot does not output duplicated code snippets in the maximum of 10 suggested code snippets. We deduce that, compared to higher-level questions, level \textbf{A} questions, which typically involve basic grammar that is simple to process, may result in a higher number of duplicated code snippets.

Comparing the average number of suggested code snippets across the three languages, we found that Japanese had the lowest number. As observed in Section 4, Japanese also leads to the highest accuracy. Therefore, there might be a relationship between the number of suggested code snippets and the accuracy.
The possible reason for this is that AtCoder is primarily operated in Japan and is widely used by Japanese people. 
Consequently, there is a larger pool of Japanese-coded responses with accompanying comments that were utilized during the training of the large language models.

\section{Threats to Validity}

\textbf{External Validity} 
The evaluation results may not be generalized to other contest questions, programming languages other than Python, and natural languages other than English, Japanese, and Chinese. 
However, we believe that the representativeness of our study is highlighted by the dominance of Python and the popularity of the natural languages chosen.
Furthermore, the questions in AtCoder are designed for learners. 
Therefore, future studies should investigate the impact of code suggestions on professional tasks.

\noindent
\textbf{Construct Validity} During the data preparation, the questions may not be translated precisely from English to Chinese~\cite{lin2022EMSE}.
To address this threat, we have included a native Chinese researcher with a strong research background in this work to manually validate the Chinese translation generated by \texttt{DeepL}.

\noindent
\textbf{Internal Validity} We relied on the suggestion tool Copilot. 
It utilizes the provided information in the file and project to construct its own internal context, but the exact details are not publicly disclosed. 

\section{Conclusion and Future Work}
This study explores the effectiveness of GitHub Copilot in the setting of different natural languages (i.e., English, Japanese, and Chinese).
Specifically, we evaluate the accuracy of the suggested code by Copilot using 
a total of 756 questions sourced from 189 AtCoder contests, with each contest providing four difficulty levels of problems.
Our results highlight that the capability varies across natural languages, with Chinese achieving the worst performance.
Additionally, regardless of the type of natural language, the performance decreases significantly as the difficulty of questions increases.

Our preliminary study has identified several anticipated future works, including:
(i) Investigating the possibility that the appropriate natural language for each programming task may differ.
(ii) Comparing the questions from other programming contests, such as Chinese-based contests.
(iii) Conducting a deeper analysis of the quality and understandability of the suggested code across languages. and (iv) explore the impact of natural languages in other state-of-the-art large-language models (such as ChatGPT).

\section*{Acknowledgments}
We gratefully acknowledge the financial support of: (1) JSPS for the KAKENHI grants (JP21H04877, JP22K17874, JP22K18630, JP23K16864), and Bilateral Program grant JPJSBP120239929; and (2) the Inamori Research Institute for Science for supporting Yasutaka Kamei via the InaRIS Fellowship.

\section*{Data Availability}
To encourage the replication study in the future, we have made our replication package available~\cite{replication}, including the experiment scripts and the complementary results.

\newpage
\balance
\bibliographystyle{ACM-Reference-Format}
\bibliography{acmart}
\end{document}